# ELECTRON EMISSION OF THE STRIPPING FOIL AND COLLIMATION SYSTEM FOR CSNS/RCS*

M.Y. Huang[#], Y.D. Liu, N. Wang, S. Wang

Institute of High Energy Physics, Beijing, China

*Abstract*

For the Rapid Cycling Synchrotron of the China Spallation Neutron Source (CSNS/RCS), the electron emission plays an important role in the accelerator limitation. The interactions between the proton beam and the stripping foil were studied, and the electron scattering processes were simulated by the ORBIT and FLUKA codes. Then, the electron energy distribution and the electron yielding rate can be given. Furthermore, the interactions between the proton beam and the collimation system were studied, and the electron scattering processes were simulated. Then, the energy distribution of the primary electron emission can be given and the yielding rate of the primary electron can be obtained.

## INTRODUCTION

CSNS is a high power proton accelerator-based facility [1]. Its accelerator consists of a 1.6GeV RCS and an 80MeV proton linac which can be upgraded to 250MeV [2]. The RCS accumulates $1.56 \times 10^{13}$ protons in two intense bunches and operates at a 25Hz repetition rate with an initial design beam power of 100kW and is capable of being upgraded to 500kW.

For high intensity proton accelerators, injection via $H^-$ stripping is a practical method [3]. The design of the RCS injection system is to inject $H^-$ beam into the RCS with high precision and high transport efficiency. In the injection system, there are two carbon stripping foils: a primary stripping foil and a secondary stripping foil. When the $H^-$ beam traverses the carbon stripping foil, most of the particles $H^-$ are converted to $H^+$, and the others are converted to $H^0$ or unchanged. In our pervious paper [4], the beam losses due to the interaction between the $H^-$ beam and stripping foil had been studied. In this paper, by using the ORBIT [5] and FLUKA [6] codes, the electron scattering processes when the proton beam traverses the carbon stripping foil will be simulated and the electron production will be studied.

In order to control the beam losses around the RCS to an acceptable level, the collimation systems are often used to localize the beam losses to restricted areas. For CSNS/RCS, much work about the collimation system had been done [7] [8]. In this paper, the interactions between the proton beam and the collimation system (one primary collimator and four secondary collimators) will be studied. By using the ORBIT and FLUKA codes, the particle scattering of the collimation system can be simulated and the energy distribution of the primary electron emission can be obtained.

___________________________________________

*Work supported by National Natural Science Foundation of China (Nos. 11205185, 11175020, 11175193)

[#]huangmy@ihep.ac.cn

## ELECTRON SCATTERING OF THE STRIPPING FOIL

The CSNS/RCS injection system aims to inject the $H^-$ beam into the RCS with high precision and high transport efficiency. When the $H^-$ beam traverses the stripping foil, 99% of the particles $H^-$ are converted to $H^+$ and two electrons per $H^-$ are produced. However, due to the interactions between the proton beam and the stripping foil, more electrons may be produced. In this section, by using the ORBIT and FLUKA codes, the electron scattering processes due to the interactions between the proton beam and the stripping foil will be simulated, and then the electron yielding rate can be obtained.

Table 1: Beam parameters of the proton distribution that hitting on the stripping foil

| Parameters | Values |
|---|---|
| $(\alpha_x, \alpha_y)$ | (0.003, 0.044) |
| $(\beta_x, \beta_y)$/m | (1.833, 4.458) |
| $(\gamma_x, \gamma_y)$/m$^{-1}$ | (0.546, 0.225) |
| $(\varepsilon_{x,99\%}, \varepsilon_{y,99\%})$/(mm·mrad) | (92, 247) |
| Average traversal number | 5 |

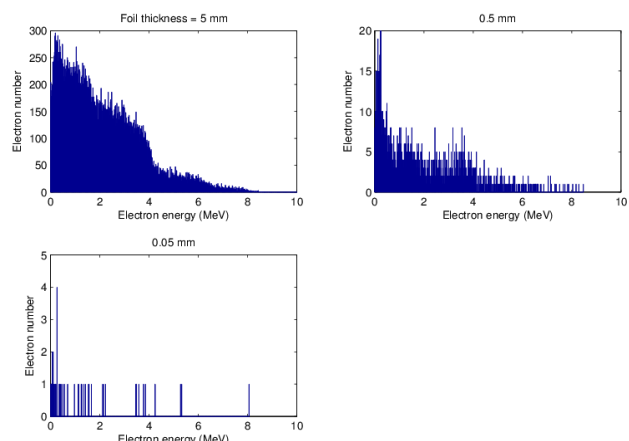

Figure 1: The electron energy distribution for the stripping foil with different thickness.

By using the code ORBIT, the injection process can be simulated. Then, the average traversal number and the beam distribution after injection can be given. Calculating those particles of the beam distribution which are in the range of the stripping foil, the twiss parameters and 99% emittance for those particles can be obtained, as shown in Table 1. With these beam parameters, the distribution of the particle beam that hitting on the stripping foil can be fitted [4]. Then, by using the FLUKA code, the electron scattering processes that the proton beam traverses the stripping foil can be simulated.

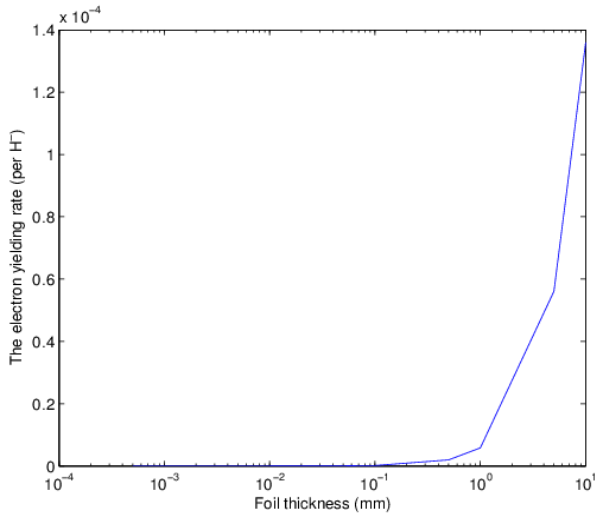

Figure 2: The electron yielding rate as a function of the foil thickness.

By using the method discussed above, the electron scattering processes are simulated when the proton beam traverses the stripping foil with different thickness. Only considering the electrons whose energy is above 1keV, Figure 1 shows different electron energy distribution for the stripping foil with different thickness and Figure 2 shows the electron yielding rate as a function of the foil thickness. It can be found that the electron yielding rate nearly equal to 0 when the foil thickness is smaller than 0.01mm. For CSNS/RCS, the thickness of the stripping foil is 100g/cm$^2$ (about $5 \times 10^{-4}$mm). Therefore, during the injection process, the additional electrons due to the interactions between the proton beam and the stripping foil can be neglected.

## PRIMARY ELECTRON EMISSION OF THE COLLIMATION SYSTEM

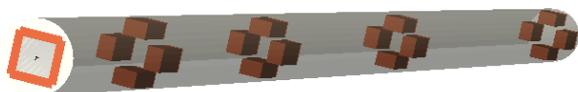

Figure 3: The model of the collimation system used by FLUKA.

For CSNS/RCS, the transverse collimation system is composed of one primary collimator and four secondary collimators. The primary collimator consists of four 0.17mm thick tungsten scrapers, which are set either horizontal or vertical. Four secondary collimators setting downstream of the primary collimator serve as absorbers. Each of the secondary collimators consists of four 200mm thick movable copper blocks. Therefore, the model of the collimation system used by the FLUKA code can be shown as Figure 3.

By using the ORBIT code, the injection process and the acceleration process can be simulated. Then, the proton beam distribution hitting on the collimation system can be obtained. By using the FLUKA code and the above proton beam distribution, the particle scattering of the collimation system can be simulated and the energy distribution of the primary electron emission can be obtained. Considering the primary electrons whose energy is above 1keV, the primary electron energy distribution can be given in Figure 4. In addition, it can be found from the simulation that the yielding rate of the primary electrons whose energy is above 1keV is about 0.1% per proton.

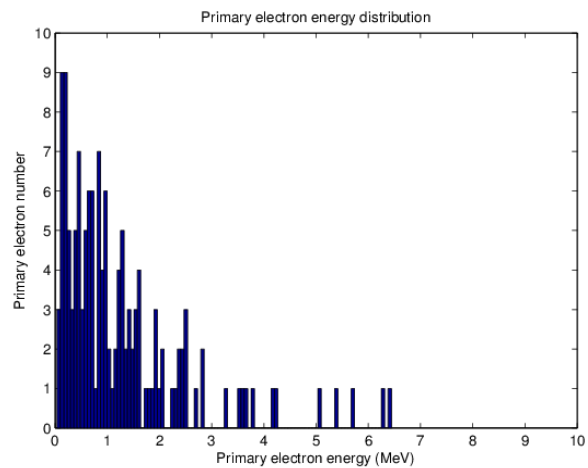

Figure 4: The primary electron energy distribution.

## CONCLUSIONS

The electron emission of the stripping foil and the collimation system for CSNS/RCS was studied in this paper. The interactions between the proton beam and the stripping foil were discussed. By using the ORBIT and FLUKA codes, the stripping foil scattering was simulated. The electron energy distribution and the electron yielding rate were obtained. It can be found that, during the injection process, the additional electrons due to the interaction between the proton beam and the stripping foil can be neglected. Secondly, the interactions between the proton beam and the collimation system were studied and the particle scattering processes in the region of the collimation system were simulated. The primary electron energy distribution and the yielding rate were obtained. It was found that the yielding rate of the primary electrons whose energy is above 1keV is about 0.1% per proton.

## ACKNOWLENDGMENTS

The authors want to thank CSNS colleagues for the discussion and consultations.